\documentclass[a4paper,12pt]{article}
\usepackage[utf8]{inputenc}
\usepackage[english]{babel}
\usepackage{amsmath, amssymb,graphicx}
\usepackage{caption}
\usepackage{subcaption}
\pdfoutput=1 
\usepackage{jheppub} 
\usepackage{amsfonts}
\usepackage{ulem}

\newcommand{\be}{\begin{equation}}
\newcommand{\ee}{\end{equation}}
\newcommand{\bea}{\begin{eqnarray}}
\newcommand{\eea}{\end{eqnarray}}
\newcommand{\nn}{\nonumber}
\newcommand{\dd}{\partial}

\newcommand{\e}{\mathrm{e}}

\title{AdS/CFT prescription for angle-deficit space and winding geodesics}

\author[a]{Irina Ya. Aref'eva,}
\author[a]{Mikhail A. Khramtsov 	}

\affiliation[a]{Steklov Mathematical Institute, Russian Academy of Sciences,\\Gubkina str. 8, 119991, Moscow, Russia}

\emailAdd{arefeva@mi.ras.ru}
\emailAdd{khramtsov@mi.ras.ru}

\abstract{
We present the holographic computation of the boundary two-point correlator using the GKPW prescription for a scalar field in the AdS$_3$ space with a conical defect. Generally speaking, a conical defect breaks conformal invariance in the dual theory, however we calculate the classical bulk-boundary propagator for a scalar field in the space with conical defect and use it to compute the two-point correlator in the boundary theory. 
We compare the obtained general expression with previous studies based on the geodesic approximation.
They are in good agreement for short correlators, and main discrepancy comes in the region of long correlations. Meanwhile, in case of $\mathbb{Z}_r$-orbifold, the GKPW result coincides with the one obtained via geodesic images prescription and with the general result for the boundary theory, which is conformal in this special case.

}

\begin{document}
\maketitle
\flushbottom
\newpage
\section{Introduction}
\label{sec:intro}
AdS/CFT  and holography  \cite{Malda,GKP,Witten,Aharony1999} have been proving to be very fruitful tools in providing a computational framework for strongly-coupled systems, as well as giving new insights into the underlying structures of string and conformal field theories. They  have demonstrated to be very useful for description  of strong interacting equilibrium and non-equilibrium system in high energy physics, in particular, heavy-ion collisions and formation of QGP \cite{Solana,IA,DeWolf}, as well as in the condensed matter physics \cite{Hartnoll08kx,Sachdev:2010ch}. The frameworks of these applications are set up essentially through consideration of different modifications of the basic AdS background, in particular, backgrounds which break asymptotic conformal symmetry of the boundary
\cite{Kanitscheider:2008kd,Gursoy:2010aa,Gursoy:2007cb,Gursoy:2007er,Buchel:2013id}. 

In the paper we consider deformations of  AdS$_3$ by conical defects. There are several reasons to consider this problem. 
First of all, AdS$_3$/CFT$_2$ allows to probe fundamental theoretical problems, such as the thermalization problem \cite{Bal11,Lopez,Kera11,Caceres:2012em,Bal12,Aref'eva:2013wma,IA15QGP}, entanglement problem and information paradoxes \cite{Ryu:2006bv,AbajoArrastia:2010yt,Callan12,Bal14}, chaos in QFT \cite{Maldacena15} using simple toy models. 
The second reason is that in this case one can distinguish the peculiar features of several approximations that are widely used in AdS/CFT correspondence. The prime example of such approximation is the holographic geodesic approximation \cite{Bal99}. It plays a very important role in holographic calculations. Many physical effects have been described within this approximation, in particular, behaviour of physical quantities such as entanglement and mutual entropies, Wilson loops during thermalization and quench are studied mainly within this approximation \cite{Hubeny10,Bal11,Bal12,Lopez,Kera11,Caceres:2012em,AbajoArrastia:2010yt,Callan12,Aref'eva:2013wma,IA15QGP,
Ryu:2006bv,Bal14,Albash:2010mv}. Recent developments in the $2D$ CFT bootstrap techniques show the deep relation between the geodesic approximation and semi-classical limit of the conformal field theory \cite{Fitz14,Alkalaev15}. 

Recently, geodesic approximation has been used extensively to study the structure of the two-dimensional CFT and its deformations which are dual to various locally AdS$_3$ backgrounds, such as BTZ black holes or Deser-Jackiw point-particle solutions. The latter is the subject of study of the present paper. The point particles in AdS$_3$ \cite{Deser, Deser2,DeserLambda,Hooft} produce conical singularities, cutting out wedges from the space, but leaving it locally AdS$_3$. We will focus on the case of the static massive particle. The recent work \cite{AB-TMF,Arefeva2015,AAT,AA} was devoted to the study of the two-point correlation function and the entanglement entropy in the boundary dual to the AdS$_3$-deficit spacetime in the framework of geodesic approximation. The main feature observed therein is  a non-trivial analytical structure of correlators, which is caused by the fact that identification of the faces of the wedge cut out by the particle allows to have, generally speaking, multiple geodesics connecting two given points at the boundary. Since this is true only for some regions of the boundary, naturally, the geodesic result for the two-point function may be discontinuous and can exhibit some peculiar behaviour in the long range region. 

 The goal of the present investigation is to study the two-point boundary correlator from the point of view of the on-shell action for the scalar field via GKPW prescription \cite{GKP,Witten} on AdS$_3$ with a conical defect, and compare the result to the one obtained from the geodesic prescription. As an interesting special case, we formulate the images prescription for the correlator in case when the space is an orbifold AdS$_3/\mathbb{Z}_r$ and compare it with the image method based on the geodesic approximation \cite{AB-TMF}. In the general case we illustrate that the discontinuities in the geodesic result correspond to the non-conformal regime. We emphasize though that since we generally deal here with conformal symmetry breaking, our study, being based on the original AdS/CFT prescription, indicates the need for caution when applying holographic methods. Although in some cases it also justifies the application of techniques based either on geodesic approximation or computation of the on-shell action, and it provides some limited evidence for a possibility of modification of AdS$_3$/CFT$_2$ prescription which could take into account non-conformal deformations of the holographic correspondence.

The paper is organized as follows. Section ~\ref{setup} contains a brief overview of the geometry of AdS$_3$ with a massive static particle in the bulk and shortly describes the Lorentzian GKPW prescription in case of the empty AdS$_3$ space. We also review the effect of the conical defect on the boundary field theory from the symmetry point of view and the geodesics prescription for deficit-angle in the bulk and its relation to the general holographic dictionary. We then proceed to generalize the GKPW approach to the case of AdS-deficit spaces in section \ref{GKPW-cone}. 
In the section \ref{integercase} we consider the special case of $\mathbb{Z}_r$-orbifold when we have a conformal theory on the boundary and compare the general result with the images prescription for geodesics. Then in section \ref{nonintegercase} we consider general non-conformal deformations in case of small and large deficit angle, as well as their effect on the temporal dependence of correlators in GKPW and geodesic prescriptions. 

\section {Setup \label{setup}}
\subsection{Scalar field on AdS$_3$ space  with particle }
We start with a brief overview of conical defects in the AdS$_3$ space. The three-dimensional geometry with a conical singularity at the origin arises as a solution of the three-dimensional Einstein gravity with a point-like source, which was obtained by Deser, Jackiw and t'Hooft originally in the flat space \cite{Deser} and generalized to the case of constant curvature in \cite{DeserLambda}. The AdS$_3$ space with a conical defect is such solution with negative cosmological constant. It represents a static massive particle sitting in the origin of the empty AdS space. This is the only place in which the particle can be at the mechanical equilibrium because any small deviation from the center get suppressed by the quadratic gravitational potential caused by the negative cosmological constant. The metric in global coordinates can be written as follows (in the present paper we set AdS radius to $1$):
\be
ds^2 = \frac{1}{\cos^2 \rho} \left(-dt^2+d\rho^2+\sin^2\rho\ d\vartheta^2 \right)\,,
\label{ads3con}
\ee
where we have $\rho \in [0, \frac{\pi}{2})$ as the holographic coordinate, AdS boundary is located at $\pi/2$; and $\vartheta \in [0, 2\pi A)$ is the angular coordinate. We parametrize the conical defect as 
\be A=1-4G\mu,
\ee where  $\mu$ is the mass of the particle, and $G$ is the three-dimensional Newtonian constant\footnote{In the case when the living space angle is $2\pi$ times an integer, i. e. when $A=s$, $s \in  \mathbb{Z}_+$, the spacetime has an angle excess. This particular case is a solitonic topological solution of the pure $3D$ gravity \cite{Izqu94}, $s$ representing the winding number.}. It is clear that the above metric indeed has the deficit angle of value \be \gamma=2\pi(1-A)=8\pi G\mu. \ee
The case of $A=0$ is the BTZ black hole threshold. 

We will consider the real scalar field on the background (\ref{ads3con}) with action \footnote{Classical and quantum theories of the scalar field   on a cone on AdS$_3$ have been considered in \cite{DeserLambda} and in the flat case 
\cite{Deser2,QFT-cone,volovich}. Recently there have been interesting developments concerning correlation functions and conformal symmetry on spaces with conical defects \cite{Bayona10,Smolkin14}.
QFT on the cone presents interest also in context of cosmic strings applications \cite{CS,CM}. }
\be
S = -\frac12\int d^3 x\sqrt{-g} \left( (\dd \phi)^2 + m^2 \phi^2 \right)\,.
\ee
The scalar equation of motion in the metric, similarly to the empty AdS case \cite{Bal98}, has the form
\be
-\ddot{\phi} + \frac{\cos^2 \rho}{\sin^2 \rho} \dd_\rho \left(\frac{\sin \rho}{\cos^2 \rho} \dd_\rho \phi \right) + \frac{1}{\sin^2 \rho} \dd_\theta^2 \phi - \frac{m^2}{\cos^2 \rho} \phi =0\,;
\label{ads3eom}
\ee
The variables are separated via the usual ansatz
\be
\phi(t, \rho, \vartheta) = \e^{i \omega t} Y(\vartheta) R(\rho)\,.
\ee

The angular dependence is determined by the one-dimensional eigenproblem for angular momentum, which factorizes from  equation (\ref{ads3eom}). Thus we have
\be
Y(\vartheta) = \e^{i \frac{l}{A}\vartheta}\,, \qquad l \in \mathbb{Z}\,; \label{angularmodes}
\ee
Substituting the ansatz into (\ref{ads3eom}), we obtain a Schroedinger-type eigenproblem  for the radial component (here the prime symbol denotes the $\rho$ derivative):
\be
-R'' - \frac{1}{\cos \rho \sin \rho} R' + \left ( \frac{l^2}{A^2 \sin^2 \rho} + \frac{m^2}{\cos^2 \rho} \right) R = \omega^2 R\,; \label{eqR}
\ee
This equation defines the bulk-boundary propagator of the scalar field, which is instrumental in construction of boundary correlators. The case of $A=1$ is the case of pure AdS$_3$, which we discuss in the following subsection. 

\subsection{The GKPW prescription for boundary correlators in global Lorentz AdS}

Our goal is to obtain the expression for a two-point correlation function of a scalar operator on the boundary of AdS$_3$ with a conical defect\footnote{The AdS/CFT correspondence for the case of presence of defects on the boundary is a subject of numerous investigations and applications, see for example \cite{Kirsch:2004km,Araujo:2015hna} .}, described by the metric (\ref{ads3con}), using the Gubser-Klebanov-Polyakov/Witten (GKPW) holographic prescription \cite{GKP,Witten}. Since we are interested in real-time correlation functions, we take the bulk (and, consequently, boundary) metric signature to be Lorentzian. To take into account a particular choice of boundary conditions for the Green's function in order to get a concrete real-time correlator (i. e. retarded, Wightman or causal), we will use the prescription in the form of Skenderis and van Rees \cite{Skenderis2008}. In the present subsection we briefly review the prescription in the case of empty AdS$_3$, i. e. $A=1$. We write 
\be
\left< \e^{i\int dt d\vartheta \ \varphi_0 \mathcal{O}} \right>_{\text{CFT}} = \e^{i S_{on-shell}[\phi]}|_{\phi|_{\text{bd}}=\varphi_0}\,;
\label{ads/cft}
\ee
where as usual, the equality is supposed to hold after renormalization.

To specify a concrete real-time two-point correlator of the operator $\mathcal{O}_\Delta$ with conformal dimension $\Delta$ obtained via functional differentiation of the CFT generating functional, we deform the contour of integration over time into a contour $C$ lying in the complex time plane. This is a generalization of imposing standard Feynman radiation boundary conditions on the path integral, which is used to get the causal correlator \cite{ASF}. The contour $C$ is deformed in such a way that it goes through the fields required by the chosen boundary conditions at $t = \pm T$ ($t$ being the parameter of the complex curve, $\pm T$ are the corner points of the contour), and the endpoints, corresponding to vacuum states in $\mathcal{Z}=\left< \Omega | \Omega\right>$ are either at imaginary infinity in the zero-temperature case, or at finite identified points, when the temperature is finite. In the current paper we consider the zero-temperature case. 

To construct the bulk dual, we deform the integration contour in the bulk on-shell action as well. As a result, we 
have the contributions from several on-shell actions: those which correspond to vertical segments are effectively Euclidean actions, and those that correspond to integration over horizontal segments, correspond to Lorentzian action. The sources $\varphi_0$ are set to zero on all Euclidean segments, and satisfy the condition $\varphi_0(\pm T, \vartheta)=0$. Thus, while the Euclidean pieces do not contribute directly into the boundary term of the on-shell action, they determine the contour in the complex frequency plane, which is used to define the bulk-boundary propagator, through the condition of smoothness of the scalar field on the contour $C$. 

The bulk-boundary propagator is defined in the boundary momentum representation as a solution $R_{\omega,l}(\rho)$ of the radial equation (\ref{eqR}) (since we consider the empty AdS case here, we set $A=1$ in this subsection), which is regular at the origin and has the leading behaviour $ R_{\omega,l}(\rho) = \varepsilon^{2h_-} + \dots$ near the boundary, where $\varepsilon=\frac{\pi}{2}-\rho$. Here we introduce a notation 
\be
h_{\pm} = \frac{1}{2}\pm \frac{1}{2}\sqrt{1+m^2}\,;
\ee
so that the $2h_+=\Delta$ corresponds to the conformal dimension of the boundary operator $\mathcal{O}_\Delta$, and $h_+ + h_-=1$ . Also, we define $\nu = h_+ - h_-$, so that $\Delta=1+\nu$. In this paper we consider only the case of $\nu \in \mathbb{Z}_+ \cup {0}$.

Because of the asymptotic definition of $R$, the solution of the Dirichlet problem for the scalar field equation in the bulk can be written as
\be
\Phi (\rho, t, \vartheta) = \frac{1}{(2\pi)^2}\sum_{l \in \mathbb{Z}} \int_\mathcal{C} d\omega\ \e^{-i\omega t+il\vartheta} \varphi_0(\omega, l) R_{\omega,l}(\rho)\,, \label{b2b}
\ee
Note, however, that in general $R$ consists of two pieces \cite{Bal98}: the non-normalizable piece with leading behaviour $\varepsilon^{2h_-}$, which grows near the boundary, and the normalizable piece with the leading behaviour 
$\alpha(\omega, l) \beta(\omega, l) \varepsilon^{2h_+}$, where
\bea
&&\alpha(\omega, l):= \frac{1}{\nu! (\nu-1)!}\frac{\Gamma(\left(h_+ + \frac12 (|l|+\omega)\right)\Gamma\left(h_+ + \frac12 (|l|-\omega)\right)}{\Gamma\left(h_- + \frac12 (|l|+\omega)\right)\Gamma\left(h_- + \frac12 (|l|-\omega)\right)}\,, \label{alpha}\\
&&\beta (\omega, l):= -\left(\psi\left(h_+ + \frac12 (|l|+\omega)\right)+ \psi\left(h_+ + \frac12 (|l|-\omega)\right) \right) + \dots\,; \label{alphabeta}
\eea
where by dots we denote the terms which are analytical in $\omega$. The digamma functions in $\beta$ are non-analytic and have poles at
\be
\label{omega}
\omega_{nl}^{\pm} = \pm(2h_+ +2n+|l|)\,,\qquad n \in \mathbb{Z}_+\cup{0}\,;
\ee
Thus normalizable modes are quantized, and while they clearly don't change the leading asymptotic near-boundary behaviour of $R$, they define the complex contour $\mathcal{C}$ in the frequency space around these poles. By adding or removing extra normalizable modes, we can deform $\mathcal{C}$ to obtain a concrete $i\epsilon$-prescription for the boundary correlator, and this is indeed happening via accounting for the smoothness conditions on the corners of the time contour $C$. 

To obtain the two-point correlator, one first obtains the one-point function, defined by
\be
\left< \mathcal{O}(t, \vartheta) \right> = \lim_{\varepsilon \to 0} i \frac{\varepsilon^{-\nu}}{\sqrt{-\eta}} \frac{\delta}{\delta \Phi(\rho, t, \vartheta)} \left[-\frac{i}{2}\int_C d^3 x \sqrt{-g}  \left( (\dd \phi)^2 + m^2 \phi^2 \right)\Big|_{\phi=\Phi}\right ]_{\text{subtr}}\,;
\ee
where all divergences are subtracted from the action, and $\eta=\tan \rho \sim 1/\varepsilon$ is the determinant of the induced metric on the slices of constant $\rho$. Note that, generally speaking, we would have also contributions from corners of the contour $C$, but they all vanish by virtue of smoothness conditions for the solution $\Phi$. The two-point correlator is then obtained by 
\be
G_\Delta(t, \vartheta; t', \vartheta') = \frac{i}{\sqrt{-\eta_0}}\frac{\delta}{\delta\varphi_0 (t', \vartheta')} \left< \mathcal{O}_\Delta(t, \vartheta) \right>\,;
\ee
where $\eta_0$ is the boundary metric determinant, which is just $1$ in our case.

Thus, for the Wightman correlator one gets
\be
\left< \mathcal{O}_\Delta(t, \vartheta) \mathcal{O}_\Delta(0, 0) \right> = \frac{ 2\nu}{\pi 
\nu! (\nu-1)!}\sum_{l \in \mathbb{Z}} \sum_{n=0}^\infty \frac{(n+\nu)!}{n!} \frac{\Gamma\left(n+|l|+\nu+1\right)}{\Gamma \left(n+|l|+1\right)} \times  \e^{-i(2h_+ +2n+|l|) (t-i \epsilon) +il\vartheta}. \label{corseries}
\ee
We can sum the series for any integer $\nu$. Note that $i\epsilon$ prescription here serves as a regulator to conduct the summation over $n$. The result for the two-point correlator of a scalar operator of dimension $\Delta = \nu+1$ is 
\be
\left< \mathcal{O}_\Delta(t, \vartheta) \mathcal{O}_\Delta(0, 0) \right> = \frac{\nu^2}{2^\nu \pi} \left(\frac{1}{\cos (t-i\epsilon)-\cos \vartheta} \right)^{\nu+1}\,. \label{W}
\ee
The $\Delta=1$ case has slightly different coefficient in front of the normalizable piece of the bulk-boundary propagator \cite{Bal98}, and the result in this case is 
\be
\left< \mathcal{O}_1(t, \vartheta) \mathcal{O}_1(0, 0) \right> = \frac{1}{\pi}\frac{1}{\cos (t-i\epsilon)-\cos \vartheta}\,. \label{Wnu=0}
\ee

Here we have reviewed the Skenderis-van Rees computation prescription for the Wightman correlator, and to obtain other real-time correlators in the integer $\Delta$ case, we can just rely on general QFT considerations. The Wightman correlator of a scalar operator of dimension $\Delta$ on a Lorentzian cylinder can be rewritten using standard Sokhotski formula trick as
\bea
G^W_\Delta (t, \vartheta) = \langle \mathcal{O}_\Delta(t, \vartheta)\mathcal{O}_\Delta(0,0)\rangle &=&  \left(\frac{1}{2(\cos (t-i\epsilon)  - \cos \vartheta)}\right)^\Delta =\label{Wightman}\\\nn &=& 
\left(\frac{1}{2\,\left|\cos t  - \cos \vartheta\right|}\right)^\Delta\,e^{-i\,\pi\,\Delta \,\cdot\,\theta(-\cos t  + \cos \vartheta)\,{\mbox {sign}}(\sin t)}\,.
\eea
If $\Delta$ is integer, we can simplify the exponential factor:
\bea
\langle \mathcal{O}_\Delta(t, \vartheta)\mathcal{O}_\Delta(0,0)\rangle &=&
\left\{
\begin{array}{ccc}
\left(\frac{1}{2\,\left|\cos t  - \cos \vartheta\right |}\right)^\Delta (-1)^{\Delta} & \,{\mbox {for} }\,\, & \cos t  - \cos \vartheta<0  \\
 \, & \,  & \  \\
  \left(\frac{1}{2\,\left|\cos t  - \cos \vartheta\right |}\right)^\Delta&  \,{\mbox {for} }\,\,  &  \cos t  - \cos \vartheta>0 
\end{array}
\right.\nn\\\nn\\
&=&\left(\frac{1}{2\,(\cos t  - \cos \vartheta)}\right)^\Delta\,. \label{Gads}
\eea
The causal Green function then reads
\be
G^c_\Delta (t, \vartheta) = \theta(t) \langle \mathcal{O}_\Delta(t, \vartheta)\mathcal{O}_\Delta(0,0)\rangle + \theta(-t) \langle \mathcal{O}_\Delta(0, 0)\mathcal{O}_\Delta(-t,\vartheta)\rangle \equiv G^W_\Delta (t, \vartheta)\,.
\ee
Thus, in the case of integer conformal dimension both Wightman and Feynman correlators are defined by the expression (\ref{Gads}), and the retarded/advanced Green's function is equal to zero.

\subsection{Boundary dual to the conical defect and AdS$_3$ orbifolds}

The theory on the boundary, which is dual to the AdS-deficit space, is a field theory on a cylinder of circumference $2\pi A$. To understand its relation to the "covering" CFT, i. e. the one dual to the empty AdS, we recall that the algebra of asymptotic symmetries, which has the Virasoro form for empty AdS, for the AdS-deficit case has to be replaced by its subalgebra, whose generators $l_n$ are defined as \cite{deBoer10,Bal14}:
\be
l_n = i A\ \e^{i n \frac{w}{A}} \dd_w \equiv A\ L_{\frac{n}{A}}\,,\;
\ee
where $w=t + \theta$. This subalgebra only has the Virasoro form as well if $A=\frac{1}{r}$, $r \in \mathbb{Z_+}$. In this case the bulk spacetime is the AdS$_3/\mathbb{Z}_r$ orbifold, and the boundary theory is a CFT with central charge $c = r \tilde{c}$ (we denote quantities from the covering CFT by tilde). Its operator algebra can be constructed from that of the covering CFT by symmetrizing operators with respect to the identification map, see \cite{Bal14} up to a normalization factor:
\be
\mathcal{O}(t, \vartheta) = \frac{1}{r}\sum_{k=0}^{r-1} \e^{i \frac{2\pi k}{r} \frac{\dd}{\dd \vartheta}} \tilde{\mathcal{O}}(t, \vartheta)\,;
\ee
This allows us to express matrix elements through those of the covering CFT as well. In particular, for a two-point correlator we have
\bea
\left< \mathcal{O}(t_1, \vartheta_1) \mathcal{O}(t_2, \vartheta_2) \right> &=& \frac{1}{r^2}\sum_{a=0}^{r-1} \sum_{b=0}^{r-1}  \e^{i \frac{2\pi a}{r} \frac{\dd}{\dd \vartheta_1}}  \e^{i \frac{2\pi b}{r} \frac{\dd}{\dd \vartheta_2}} \langle \tilde{\mathcal{O}}(t_1, \vartheta_1) \tilde{\mathcal{O}}(t_2, \vartheta_2) \rangle \nn\\ &=& \frac{1}{r^2}\sum_{a=0}^{r-1} \sum_{b=0}^{r-1}   \langle \tilde{\mathcal{O}}\big(t_1, \vartheta_1+ \frac{2\pi a}{r} \big) \tilde{\mathcal{O}}\big(t_2, \vartheta_2+\frac{2\pi b}{r}\big) \rangle \nn\\&=& \frac{1}{r^2}\sum_{a=0}^{r-1} \sum_{b=0}^{r-1}   \langle \tilde{\mathcal{O}}\big(t_1, \vartheta_1+ \frac{2\pi (a-b)}{r}\big) \tilde{\mathcal{O}}(t_2, \vartheta_2) \rangle \nn\\&=& \frac{1}{r}\sum_{k=0}^{r-1}  \langle \tilde{\mathcal{O}}\big(t_1, \vartheta_1+ \frac{2\pi k}{r}\big) \tilde{\mathcal{O}}(t_2, \vartheta_2) \rangle \,. \label{corCFTorb}
\eea
Hence we've obtained the expression for the correlator as a sum over images, which is what we expect for orbifold-like spaces\footnote{The similar known applications of the images method other than the AdS$_3/\mathbb{Z}_r$ orbifold case are thermal AdS case \cite{Skenderis2008}, the BTZ black hole case \cite{Bytsenko97,Skenderis2008} and multi-boundary AdS orbifold constructions \cite{Bal03}.}. For general $A$ we emphasize that the boundary algebra of symmetries does not have Virasoro form, and thus the theory is not conformally invariant. As we will demonstrate, this can be seen directly from the holographic expression for the two-point function obtained from geodesic approximation. 

\subsection{Extrapolation BDHM dictionary and geodesics approximation \label{geodesicApp}} 

Since we are also interested in comparison of the GKPW prescription and geodesic approximation for purposes of calculation of two-point functions in the AdS-deficit space, it will be useful to briefly recall the general relation between these prescriptions and the results given by the geodesics prescription. 

Originally \cite{Bal99}, the geodesic prescription was suggested as an approximation to the boundary propagator in the Euclidean AdS space, which is obtained from the bulk propagator using the dictionary which extrapolates the bulk fields $\Phi(\rho, t, \vartheta)$ to the boundary, or BDHM dictionary \cite{Banks1998}. In coordinates (\ref{ads3con}), this is expressed in defining boundary fields as
\be 
\mathcal{O}_\Delta(t, \vartheta) = \lim_{\rho \to \frac{\pi}{2}} (\cos \rho)^{-\Delta} \Phi (\rho, t, \vartheta)\,.
\ee
The bulk propagator for a scalar field can be written in the worldline representation as a path integral over particle trajectories and approximated using the leading order of the steepest descent expansion:
\be 
G_{\text{bulk}}(A, B) = \int_A^B \mathcal{D} \mathcal{P} \e^{-m \int d\lambda \sqrt{\dot{x}^2}} \sim \sum_{\text{saddles}} \e^{-m \mathcal{L}(A, B)}\,, \label{path}
\ee
where $\mathcal{D}\mathcal{P}$ is the measure on the space of particle trajectories between  points $A$ and $B$ which includes the Faddeev-Popov determinant originating from the worldline reparametrization invariance, $\lambda$ is the parameter of a trajectory and $\mathcal{L}(A, B)$ is the geodesic length between $A$ and $B$. It is implied that $m \sim \Delta$ is large. The BDHM dictionary then leads to consideration of geodesics between two boundary points, with divergences subtracted from their lengths.

It was conjectured in \cite{Banks1998} and proven in \cite{Harlow2011} that the BDHM dictionary in case of (locally) asymptotically AdS spacetime is equivalent to the GKPW dictionary. Therefore, in our case we can consider the geodesic boundary correlator as the approximation to the full GKPW expression, i. e. in Euclidean case
\be 
G_{\text{GKPW}} = G_{\text{BDHM}} \sim \sum_{\text{saddles}} \e^{-m \mathcal{L}}\,. \label{GKPW_BDHM}
\ee
Thus the geodesic approximation is given in the leading order of the WKB approximation to the full GKPW expression. 

One can formulate the geodesics prescription in the Lorentzian space, that would be valid for points $A$ and $B$ that are spacelike separated. However, for timelike separated boundary points there are no connecting geodesics. If the spacetime allows for Euclidean analytic continuation, one can obtain the Lorentzian correlator which would be valid for the entire Lorentzian plane of the boundary by making the reverse transition from the Euclidean case. However, for more general backgrounds (for example non-stationary ones) one does not have this opportunity, so for timelike separated points the geodesics prescription has to be formulated using different considerations. As an example, below we briefly discuss possible continuation into the timelike region for geodesic prescription on the background of the moving particle. 
\subsection{Geodesics image method for AdS-deficit spacetime} 

The geodesic prescription for particles in AdS$_3$ has been considered in \cite{Arefeva2015,AB-TMF,AAT,AA,AKT}. it is based around the fact that there can be several geodesics between two points in general, which differ in number of windings around the defect. It is proven that the lengths of winding geodesics can be expressed through lengths of geodesics connecting certain auxiliary points at the boundary. These points are images of the correlator arguments with respect to the isometry corresponding to the identification of faces of the wedge. The prescription is formulated in the Lorentzian signature. Thankfully, in the case of the static AdS-deficit spacetime, there is a straightforward Euclidean analytic continuation, so one can obtain the Lorentzian geodesics prescription for the entire boundary spacetime by making the reverse Wick rotation from the Euclidean case. For the Wightman correlator, that is expressed in the transition $\tau \to it +\epsilon$. The resulting correlator obtained via the geodesics prescription on the conical defect is  written as a sum over images. 
\begin{itemize}
\item For small deficit, $\frac12<A<1$, the correlator is:
\bea
G^W_\Delta(t,\vartheta)=\theta(\pi - \vartheta)\left(\frac{1}{2(\cos (t-i\epsilon)- \cos \vartheta)}\right)^{\nu+1}+\\\nn \theta (\vartheta-\pi+\gamma) \left(\frac{1}{2(\cos (t-i\epsilon)- \cos (\vartheta+\gamma))}\right)^{\nu+1} \label{smallgeod}\,;
\eea
\item For large deficit $0<A\leq\frac12$, the correlator is given by 
\bea
G^W_\Delta(t, \vartheta) =&&\left(\frac{1}{2(\cos (t-i\epsilon) -\cos \vartheta)}\right)^{\nu+1}  + \sum_{k=1}^{k_{max}} \left(\frac{1}{2(\cos (t-i\epsilon) -\cos (\vartheta+2\pi A k))}\right)^{\nu+1} \nn\\
&&+\sum_{j=1}^{j_{max}} \left(\frac{1}{2(\cos (t-i\epsilon) -\cos (\vartheta-2\pi A j))}\right)^{\nu+1}\!\!\!, \label{geodesicsImages}
\eea
where (square brackets represent the integer part):
\be
k_{max} = \left[\frac{\pi - \vartheta}{2\pi A}\right]\,, \qquad j_{max} = \left[\frac{\pi + \vartheta}{2\pi A}\right]\,; \label{geodesicsNumber}
\ee
The angular dependence in theta-functions and in limits of summation says that these functions are generally discontinuous. We will discuss the analytic structure of the geodesic correlators more thoroughly when we compare them to the result of GKPW calculation.
\end{itemize}

In the general case of moving particle, there is no straightforward way to perform Euclidean analytic continuation, and one has to formulate a separate prescription for the timelike region, for example based on quasigeodesics method \cite{Bal12,AAT} and symmetry properties \cite{AAT,AKT}. The quasigeodesics method accurately captures the behaviour of the pre-exponential factor in the expressions (\ref{Wightman}) for every term in the sum over images in the timelike region. It yields that the correlator for timelike separated points $(t_1, \vartheta_1)$ and $(t_2, \vartheta_2)$ in empty AdS can be computed using the spacelike geodesic between the points $(t_1+\pi, \vartheta_1+\pi)$ and $(t_2, \vartheta_2)$ \cite{AKT}. 
 Keeping it in mind, for integer conformal weights one can formulate the geodesic images prescription for the entire correlator on the AdS-deficit space taking into account its causal structure, without relying on the Euclidean analytic continuation. As seen in (\ref{Gads}), in the case of integer $\Delta$, the expressions for correlation functions are significantly simplified.

\section{GKPW prescription  for AdS$_3$ with static particles \label{GKPW-cone}}

Now we consider the scalar field equation in the space with metric (\ref{ads3con}) for arbitrary $A \in (0,\ 1)$. It is clear that the form of the equation is the same as in the case of empty AdS. The only difference is that now the angular eigenfunctions are defined by (\ref{angularmodes}). Therefore, the radial wave equation is the same as in pure AdS$_3$, only with $l$ divided by $A$. Consequently, the general solution of the scalar EOM on the angle-deficit AdS$_3$ space is obtained from that on pure AdS$_3$ by transition $l \to l/A$. Tracing this replacement through the GKPW computation scheme outlined above, we infer that it will lead to the change of location of poles of digamma functions, which now are at
\be
\tilde{\omega}_{nl}^{\pm} = \pm(2(h_+ +n)+\frac{|l|}{A})\,,\qquad n \in \mathbb{Z}_+\cup {0}  \,; \label{polescon}
\ee
Expressions (\ref{alpha}-\ref{alphabeta}) now read 
\bea
&&\alpha(\omega, l):= \frac{1}{\nu! (\nu-1)!}\frac{\Gamma(\left(h_+ + \frac12 \left(\frac{|l|}{A}+\omega\right)\right)\Gamma\left(h_+ + \frac12 \left(\frac{|l|}{A}-\omega\right)\right)}{\Gamma\left(h_- + \frac12 \left(\frac{|l|}{A}+\omega\right)\right)\Gamma\left(h_- + \frac12 \left(\frac{|l|}{A}-\omega\right)\right)}\,, \label{alphacone}\\
&&\beta (\omega, l):= -\left(\psi\left(h_+ + \frac12 \left(\frac{|l|}{A}+\omega\right)\right)+ \psi\left(h_+ + \frac12\left(\frac{|l|}{A}-\omega\right)\right) \right) + \dots\,; \label{alphabetacone}
\eea 
Therefore, the resulting expression in the form of series over residues in the frequency space for the Wightman two-point function will now read
\bea
&&\left< \mathcal{O}_{1+\nu}(t, \vartheta) \mathcal{O}_{1+\nu}(0, 0) \right> = \frac{2\nu}{\pi}\sum_{l \in \mathbb{Z}} \sum_{n=0}^\infty \alpha(\tilde{\omega}_{nl}, |l|)\e^{-i\tilde{\omega}^+_{nl} t +i\frac{l}{A}\vartheta} =\\&&\nn= \frac{2}{\pi (\nu-1)^2!}\sum_{l \in \mathbb{Z}} \sum_{n=0}^\infty \frac{(-1)^\nu (n+\nu)!}{n!} \frac{\Gamma\left(-n-\frac{|l|}{A}\right)}{\Gamma \left(-n-\frac{|l|}{A}-\nu\right)} \e^{-i\tilde{\omega}^+_{nl} (t-i\epsilon) +i\frac{l}{A}\vartheta}\,.
\eea
In the form analogous to the (\ref{corseries}) the resulting expression is:
\be
\left< \mathcal{O}_{1+\nu}(t, \vartheta) \mathcal{O}_{1+\nu}(0, 0) \right> = \frac{2}{\pi (\nu-1)!^2}\sum_{l \in \mathbb{Z}} \sum_{n=0}^\infty \frac{(n+\nu)!}{n!} \frac{\left(n+\frac{|l|}{A}+\nu\right)!}{\left(n+\frac{|l|}{A}\right)!} \times \e^{-i(1+\nu+2n)t -i\frac{|l|}{A} t +i \frac{l}{A} \vartheta} \,; \label{Wcon}
\ee
where we have omitted the $\epsilon$-prescription. We can sum the series for $\nu=0$, which gives the result for $\Delta=1$:
\be
\!\!\!\!\!\!\!\!\!\!\left< \mathcal{O}_1(t, \vartheta) \mathcal{O}_1(0, 0) \right> = \frac{1}{\pi} \, \frac{\sin  \frac{t}{A}}{\sin t} \,\frac{1}{\cos \frac{t}{A} - \cos\frac{\vartheta}{A}}\,. \label{corrcyl}
\ee
Thus, the result for arbitrary integer $\nu=\Delta-1$ can be obtained using the differentiation under the sum and formally written as
\be
\!\!\!\!\!\!\!\!\!\!\left< \mathcal{O}_{1+\nu}(t, \vartheta) \mathcal{O}_{1+\nu}(0, 0) \right> = \frac{\nu}{2^{\nu} (\nu-1) ! \pi} (-1)^\nu \frac{\dd^\nu}{\dd (\cos t)^\nu}\left( \frac{\sin  \frac{t}{A}}{\sin t} \frac{1}{\cos \frac{t}{A} - \cos\frac{\vartheta}{A}}\right)\,. \label{corrcylnu}
\ee
\section {Comparison of GKPW prescription for AdS$_3$-cone
with geodesic image method. Integer $1/A$ case \label{integercase}}
Consider the case when $2\pi$ is an integer number of the angle deficits, i. e. $A=1/r$ and $r $ is an integer, and the space is the AdS$_3/\mathbb{Z}_r$ orbifold. We have from the general formula (\ref{Wcon}):
\be
\left< \mathcal{O}_{1+\nu}(t, \vartheta) \mathcal{O}_{1+\nu}(0, 0) \right> = \frac{2}{\pi (\nu-1)!^2}\sum_{l \in \mathbb{Z}} \sum_{n=0}^\infty \frac{(n+\nu)!}{n!} \frac{(n+r|l|+\nu)!}{(n+r|l|)!} \times \e^{-i(1+\nu+2n)t -i|l|r t +i l r \vartheta}\,.
\ee
Using the  identity 
\bea \label{identity}
&\,&\frac{2}{\pi (\nu-1)!^2}\,\sum_{l \in \mathbb{Z}} \sum_{n=0}^\infty \frac{(n+\nu)!}{n!} \frac{(n+r|l|+\nu)!}{(n+r|l|)!} \times \e^{-i(1+\nu+2n)t -i|l|r t +i l r \vartheta}\\
&=& \frac{\nu^2}{2^{\nu} r\ \pi}\sum_{k=0}^{r-1} \left(\frac{1}{\cos t- \cos \left(\vartheta+2\pi \frac{k}{r}\right)}\right)^{\nu+1}\,.\nn
\eea
we get for arbitrary integer $\Delta>1$
\bea\label{images}
\left< \mathcal{O}_{\Delta}(t, \vartheta) \mathcal{O}_{\Delta}(0, 0) \right> &=&
 \frac{2(\Delta-1)^2}{ \pi\,r}\sum_{k=0}^{r-1} \left(\frac{1}{2(\cos t- \cos \left(\vartheta+2\pi \frac{k}{r}\right))}\right)^{\Delta}\,. 
\eea
For the special case $\Delta=1$ one can analogously obtain
\bea\label{images1}
\left< \mathcal{O}_1(t, \vartheta) \mathcal{O}_1(0, 0) \right> &=&
 \frac{1}{\pi r}\,\sum_{k=0}^{r-1} \,\frac{1}{\cos t- \cos \left(\vartheta+2\pi \frac{k}{r}\right)}\,.
 \eea

To prove (\ref{identity}), consider the sum over $l$:
\bea 
&&\sum_{l \in \mathbb{Z}} \frac{(n+r|l|+\nu)!}{(n+r|l|)!} \times \e^{-i|l|r t +i l r \vartheta}\\\nn &=&  \sum_{l=-\infty}^{\infty}  \frac{(n+r|l|+\nu)!}{(n+r|l|)!} \frac{1}{r} \sum_{p=0}^{r-1} \e^{-i|l|r t +i r l \left(\vartheta+\frac{2\pi p}{r}\right)} \\& = & \sum_{l=-\infty}^{\infty}  \frac{(n+|l|+\nu)!}{(n+|l|)!} \frac{1}{r} \sum_{p=0}^{r-1} \e^{-i|l| t +i  l \left(\vartheta+\frac{2\pi p}{r}\right)}\nn \\&-& \sum_{q=1}^{r-1}  \sum_{l=-\infty}^{\infty}  \frac{(n+r|l|+q+\nu)!}{(n+r|l|+q)!} \frac{1}{r} \sum_{p=0}^{r-1} \e^{-i(|l|r+q) t +i (r l+q) \left(\vartheta+\frac{2\pi p}{r}\right)}\nn
\eea
The summation over $p$ in the last term can be conducted:
\bea
&&\sum_{p=0}^{r-1} \e^{\frac{2\pi p q}{r}} \e^{2\pi l p} = \sum_{p=0}^{r-1} \e^{\frac{2\pi p q}{r}} = \frac{1-\e^{2\pi i q }}{1-\e^{\frac{2 \pi i q}{r}}} = 0 \quad \forall q  \label{sumq}\,;
\eea
Therefore, the entire $q$-sum vanishes, and we have 
\bea
&\,&\left< \mathcal{O}_{1+\nu}(t, \vartheta) \mathcal{O}_{1+\nu}(0, 0) \right> =\\&=&\nn \frac{1}{r} \sum_{k=0}^{r-1}\frac{2}{\pi (\nu-1)!^2}\sum_{l \in \mathbb{Z}} \sum_{n=0}^\infty \frac{(n+\nu)!}{n!} \frac{(n+|l|+\nu)!}{(n+|l|)!} \times \e^{-i(1+\nu+2n)t -i|l| t}\e^{i l \left(\vartheta+\frac{2\pi k}{r}\right)} \,,
\eea
which, in analogy to the empty AdS result (\ref{corseries}), is precisely the sum over images (\ref{images}). 
A particular case of the formula (\ref{images}) was obtained in \cite{Bal05} in case of a massless scalar field (i. e. $\Delta=2$) by using the images prescription for the bulk-boundary propagator itself. 

The answer for geodesic correlator in the orbifold case is given by the formula (\ref{geodesicsImages}), where we set $A = \frac{1}{r}$:
\be
\left< \mathcal{O}_\Delta(t, \vartheta) \mathcal{O}_\Delta(0, 0) \right> \sim \sum_{k=0}^{r-1}\, \left(\frac{1}{2\left(\cos t- \cos \left(\vartheta+2\pi \frac{k}{r}\right)\right)}\right)^{\Delta}\,. \label{geodesicsorbifold}
\ee
The normalization factor dependent on the conformal dimension is scheme dependent and is not reproduced by the geodesic approximation, however the GKPW result (\ref{images}) has a factor $1/r$ as well, which generally does not come from a saddle point expansion. However, it is required from the point of the boundary CFT, as seen in (\ref{corCFTorb}). 

Thus, the two-point correlator on the boundary CFT dual to the $\text{AdS}_3/\mathbb{Z}_r$ orbifold is precisely reproduced by the GKPW prescription, and also by the geodesic approximation up to a numerical factor. 

\section {Comparison of GKPW prescription for AdS$_3$-cone
with geodesic image method. Non-integer $1/A$ case \label{nonintegercase}}

In this case there is no obvious way of rewriting the sum (\ref{Wcon}) in terms of the geodesic contributions. 
We are going to compare it with the geodesic result in some special cases. Before we proceed, note that since the geodesic prescription does not fix the overall numerical factor, we have to choose it manually. In the orbifold case we have seen that the GKPW result gives an extra factor of $\frac{1}{r}=A$, so it is natural for us to propose the normalization for the geodesic correlator equal to $\frac{2 \nu^2}{\pi} A$. 

 \begin{figure}[t]
\centering
\includegraphics[width=6cm]{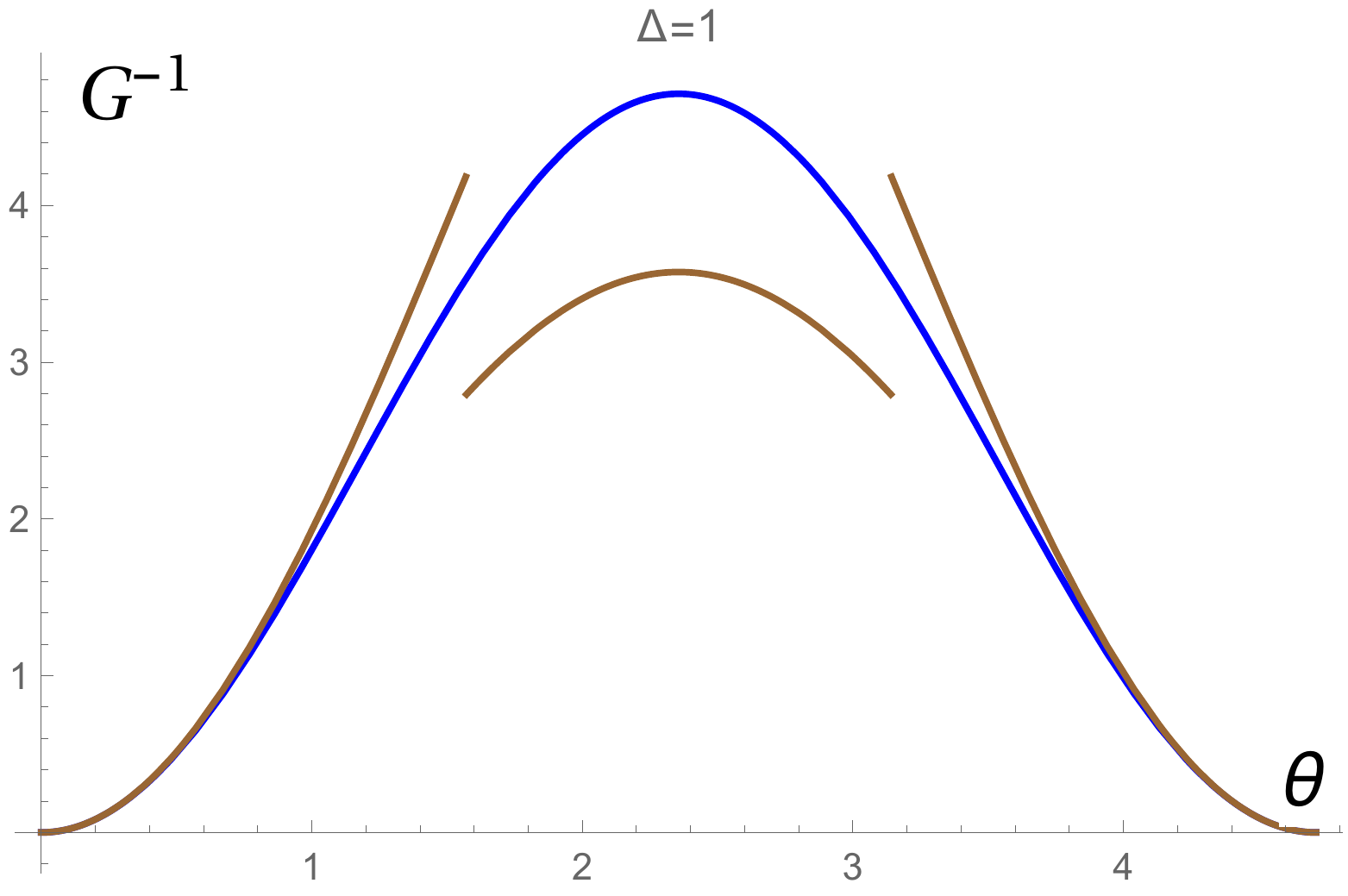}A.
\includegraphics[width=6cm]{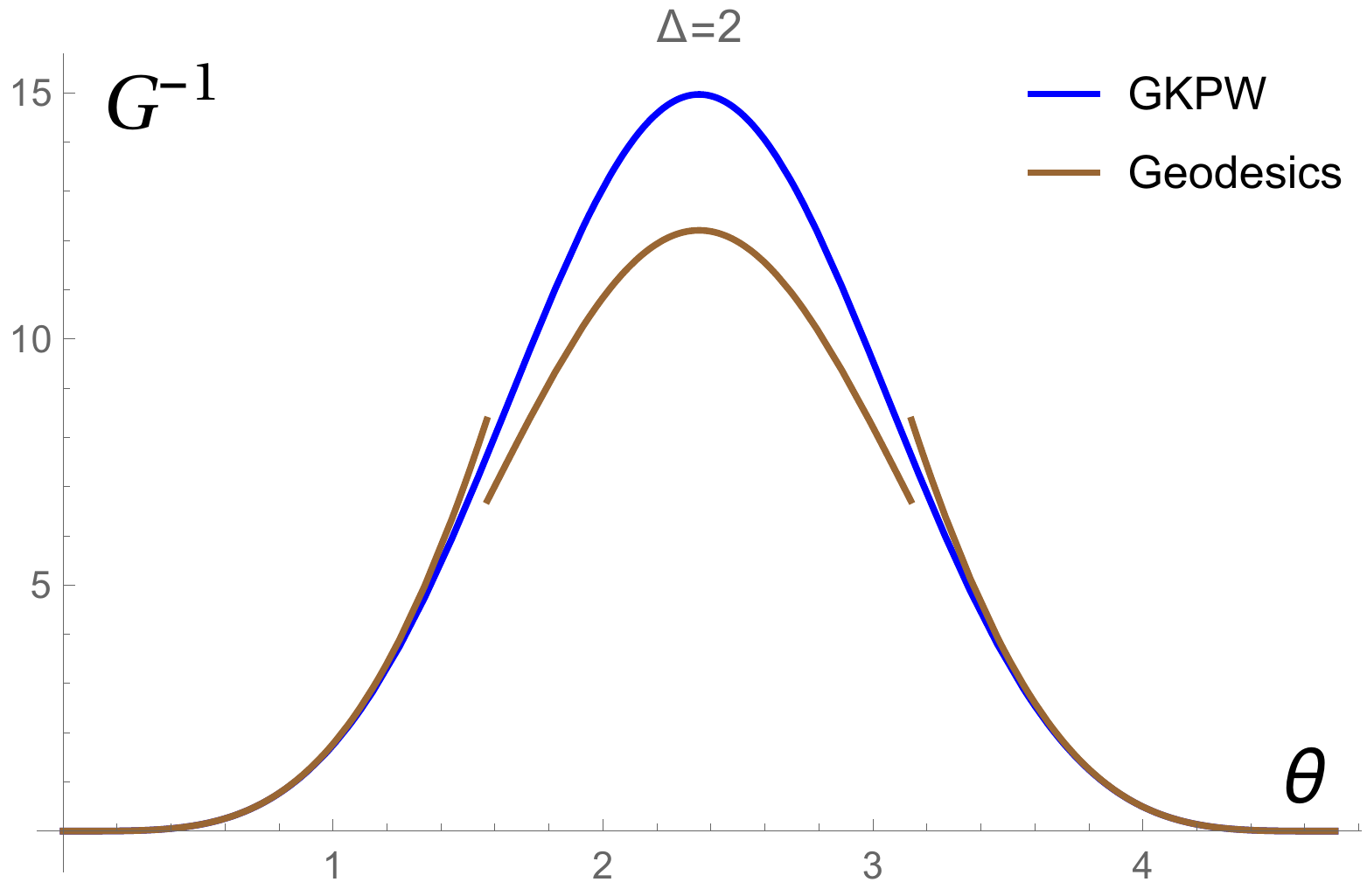}B.\\
$\,\,\,\,\,\,\,\,\,,\,\,\,$\\
\includegraphics[width=6cm]{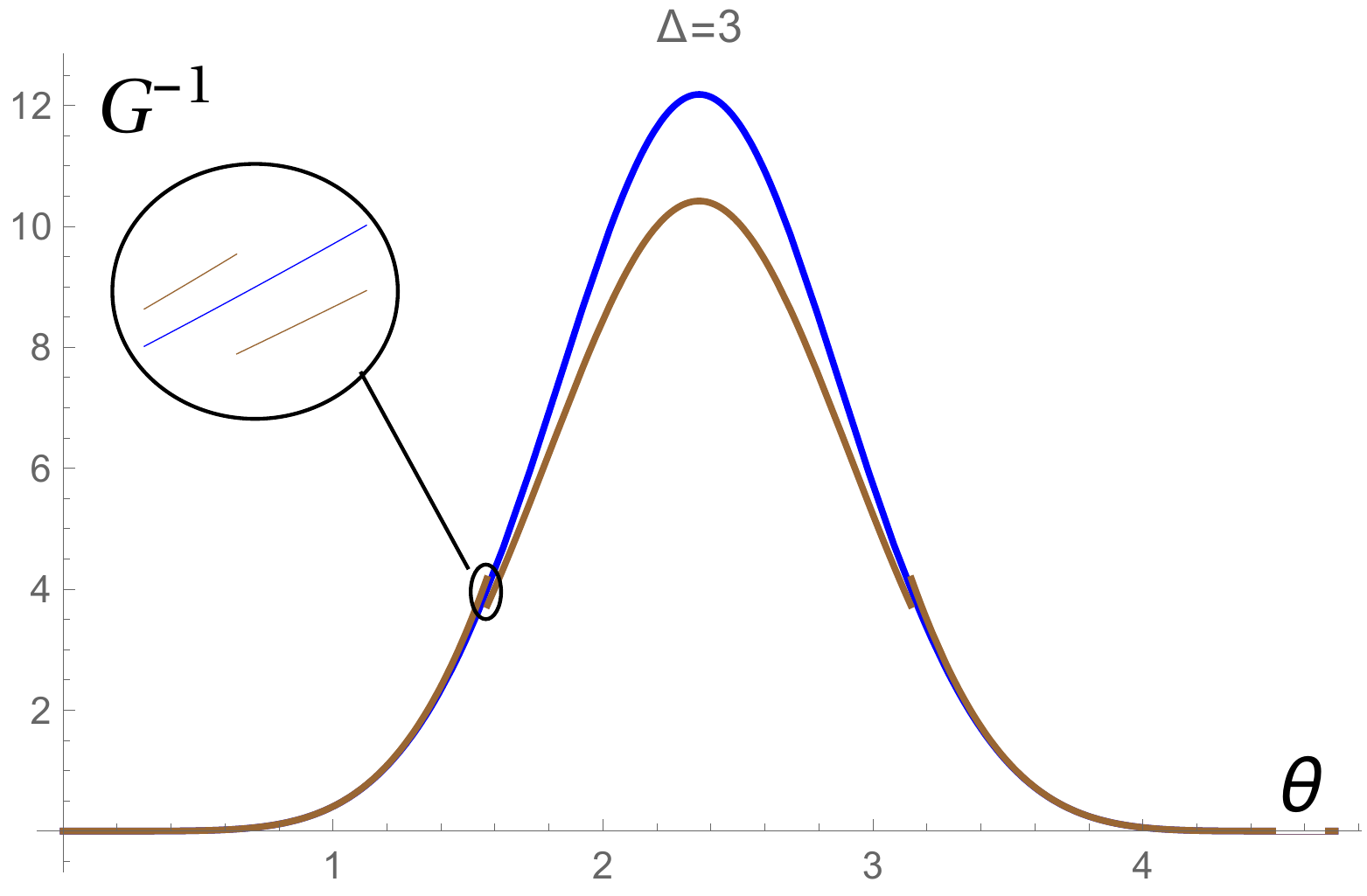}C.
\includegraphics[width=6cm]{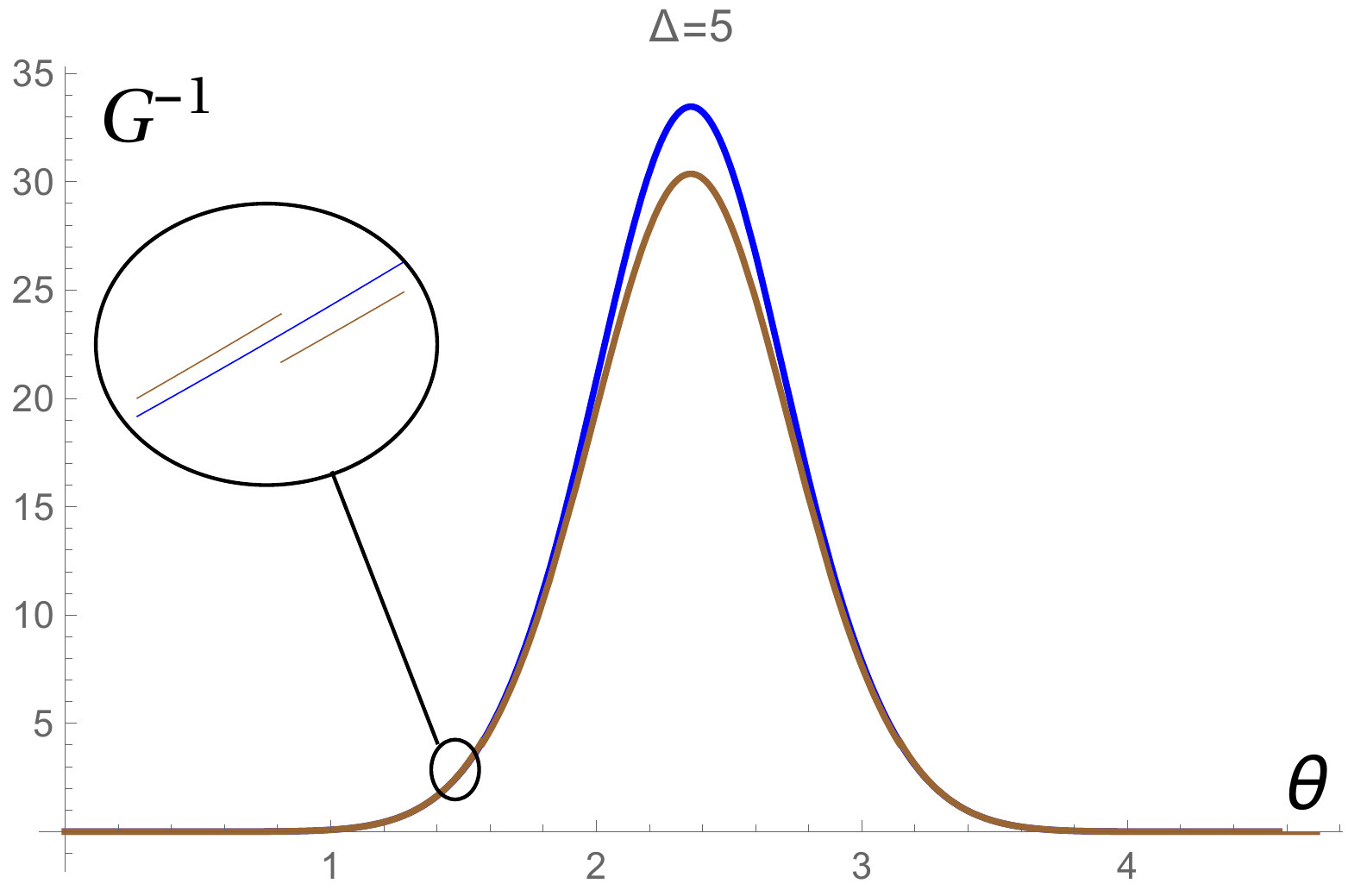}D.
\caption{
Inverse equal time correlators obtained via GKPW prescription and the geodesic image method for $A=\frac34$ for different conformal weights. Contributions of discontinuities in the geodesic result (represented by the brown line) diminish as $\Delta$ increases.}\label{Fig:nu0-5small}
\end{figure} 
\subsection{Equal time correlators}
\subsubsection{Small deficit}
Here by "small" we mean that $\gamma<\pi$. In this case there is always a region where $2$ geodesics contribute instead of only one in the remaining part of the living space. In this case the geodesic approximation predicts the correlator in the form (\ref{smallgeod}), which for convenience we rewrite as (recall that $\gamma = 2\pi(1-A)$ is the angle removed by the defect):
\bea
G(t, \vartheta) &=& \frac{2 \nu^2}{\pi} A\left[\theta (\vartheta-\pi ) \left(\frac{1}{2 \left(\cos t-\cos \left(\gamma+\vartheta\right)\right)}\right)^{\nu+1}\right.\label{largeAgeodesic}\\\nn &+&\left. \left(\frac{1}{2 (\cos t-\cos (\vartheta))}\right)^{\nu+1}\!\!\!\! \theta \left(\left(\pi -2\gamma\right)-\vartheta\right) \right.\\\nn &+&\left. \theta (\pi -\vartheta) \theta \left(\vartheta-\left(\pi -\gamma\right)\right) \left(\left(\frac{1}{2 \left(\cos t-\cos \left(\gamma+\vartheta\right)\right)}\right)^{\nu+1}\!\!\!\!+\left(\frac{1}{2 (\cos t-\cos \vartheta)}\right)^{\nu+1}\right)\right]\,.
\eea 
There are three zones: 
\begin{itemize}
\item $\vartheta \in [0, \pi - \gamma)$: The only contribution is the direct geodesic from $0$ to $\vartheta$.
\item $\vartheta \in (\pi, 2\pi - \gamma]$: The only contribution is the image geodesic from $2\pi-\gamma$ to $\vartheta$.
\item  $\vartheta \in (\pi - \gamma, \pi)$: Both direct and image geodesics contribute.
\end{itemize}
At the endpoints of these intervals we have discontinuities, which are reflected by Heaviside functions in the above formula. However, the general GKPW result (\ref{Wcon}) does not have these discontinuities. We can observe that for higher $\nu$ the size of discontinuities diminishes, and at $\Delta \to \infty$ the geodesic result approaches the GKPW expression. Examples, illustrating this point, are presented in Fig.\ref{Fig:nu0-5small}. 
Note that in the small deficit case the GKPW value is between two values of the geodesic correlator at points of discontinuity.
 \begin{figure}[t]
\centering
\includegraphics[width=6cm]{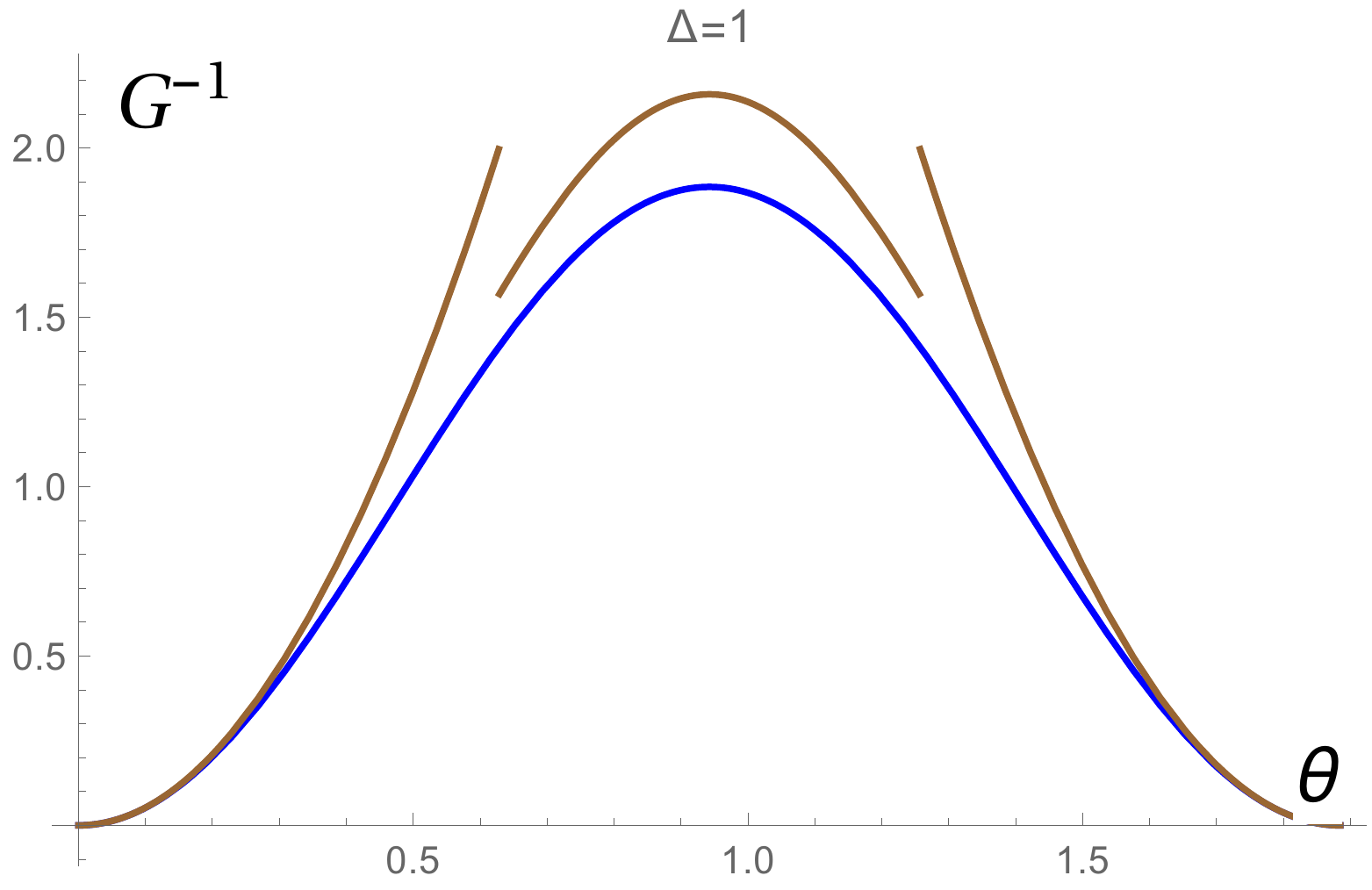}A.
\includegraphics[width=6cm]{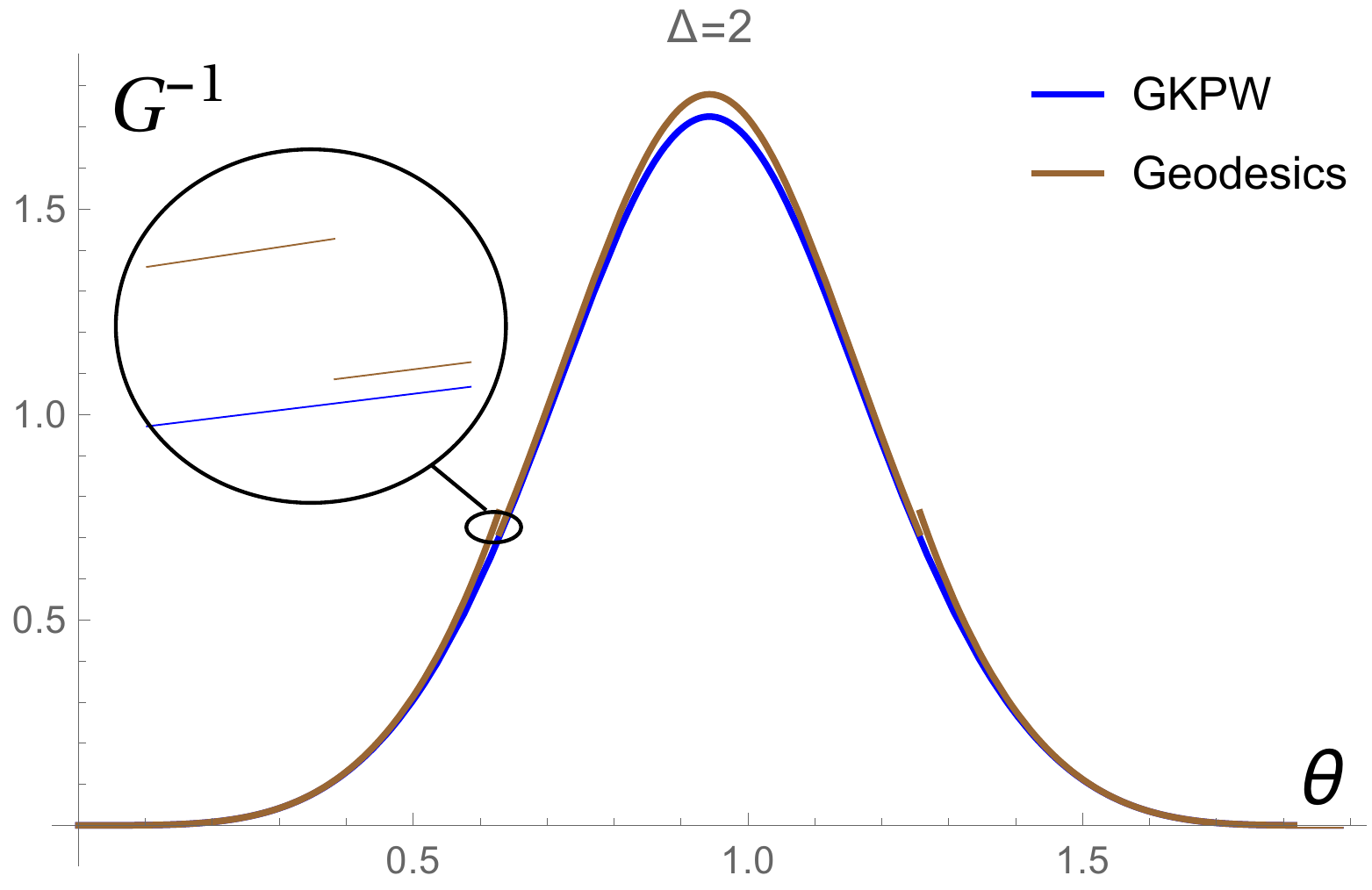}B.\\
$\,\,\,\,\,\,\,\,\,,\,\,\,$\\
\includegraphics[width=6cm]{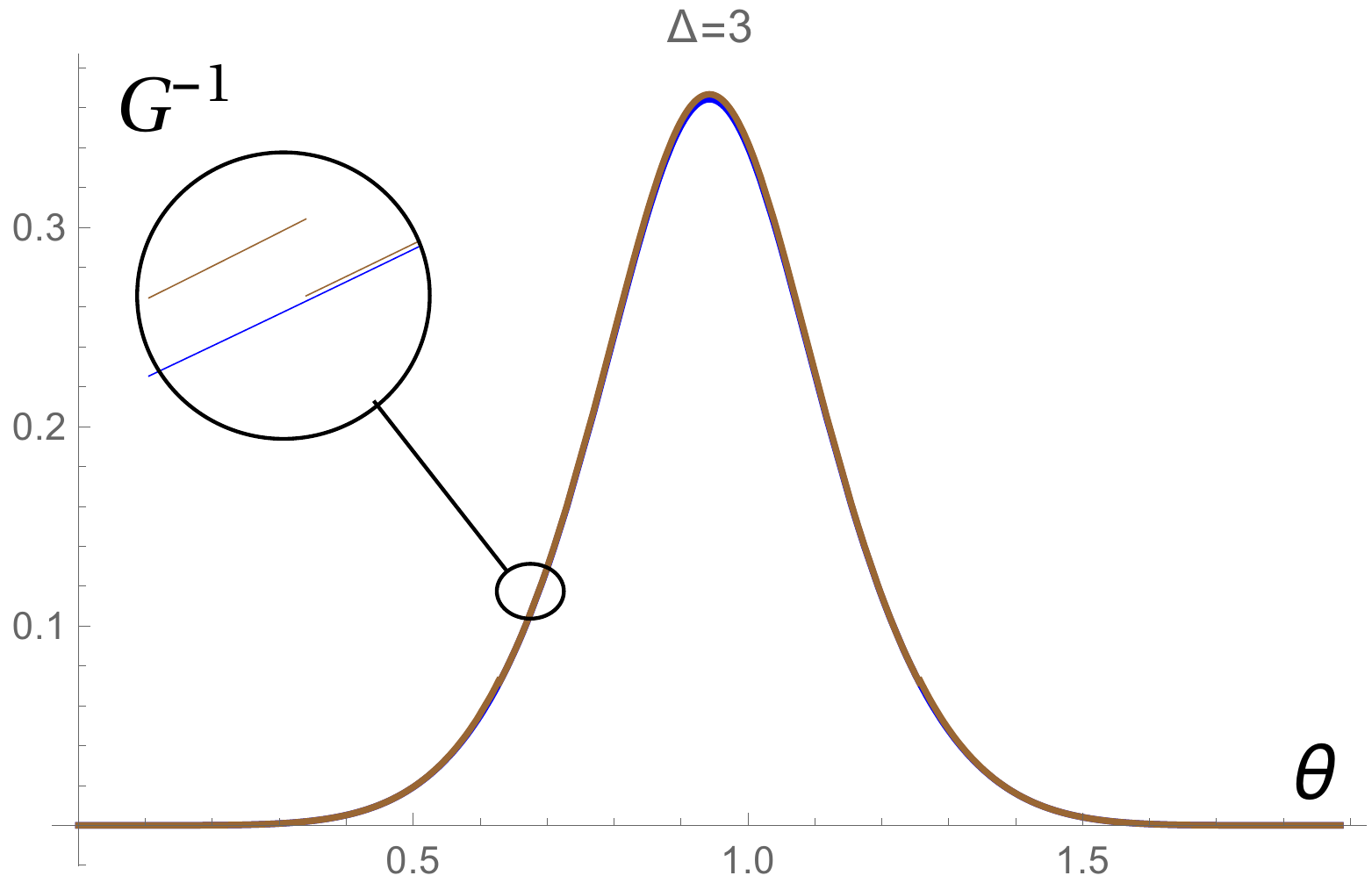}C.
\includegraphics[width=6cm]{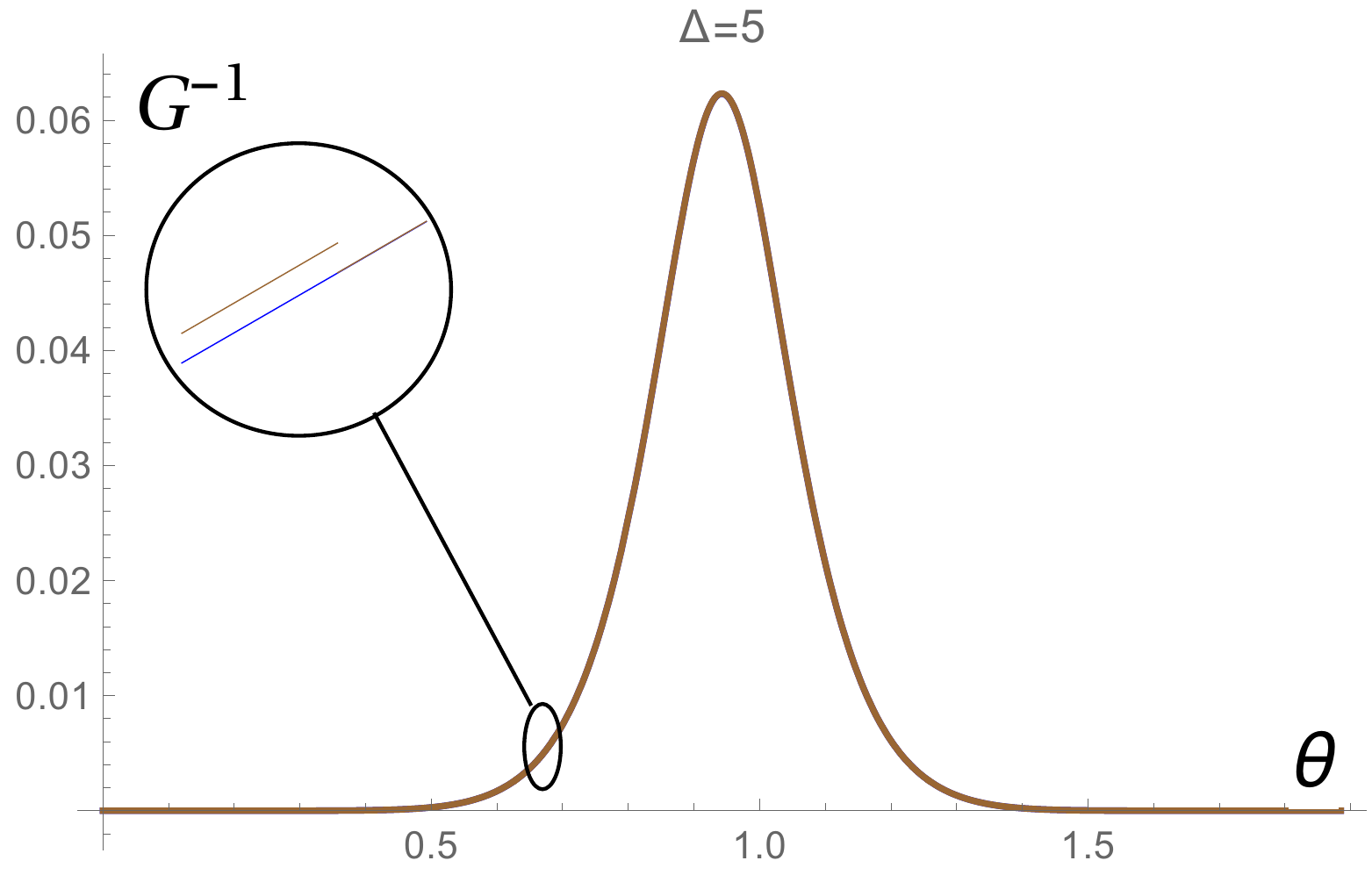}D.
\caption{
Equal time correlators obtained via GKPW prescription and the geodesic image method for $A=0.3$. The living space angle in this case equals $\frac{3 \pi}{5}$, which cannot fit into $2\pi$ integer number of times, so we have discontinuities in the geodesic result, which bring significant discrepancy with the GKPW result. However, this discrepancy also diminishes as we increase the conformal weight.} \label{Fig:nu0-5large}
\end{figure} 

Also, we see that the most significant discrepancy happens in the zone of longest correlations, which suggests that geodesic approximation apparently obtains some subleading corrections which are prominent in the the long-range correlations region. A similar effect was observed in \cite{Lin15} in the context of Vaidya model for thermalization. We leave the issue of long-range corrections in the geodesic approximation for the future study. 
\subsubsection{Large deficit}
In the case when the deficit angle is more than $\pi$, or equivalently when $A< 1/2$, the geodesic correlator is given by (\ref{geodesicsImages}). For now we focus on the case of equal-time geodesics. Because of the angular dependence in the limits of summation, the equal time correlator in the large angle case has three zones as well - transition from one zone to another corresponds to the change in the number of geodesic images given by (\ref{geodesicsNumber}). These constraints come from the fact that spacelike basic geodesics connecting two points at the boundary of AdS cannot revolve around the origin. Two sums express the two sets of images obtained by acting with the wedge identification isometry like a rotation clockwise or counter-clockwise with respect to the origin in the equal-time section of AdS cylinder. Naturally, depending on the position of the correlator endpoints and the value of the deficit angle, the number of clockwise and counter-clockwise images will change, which is what expressed by (\ref{geodesicsNumber}). In the orbifold case discussed in \ref{integercase} the identification isometry  is a $\mathbb{Z}_r$ group of rotations, so any image can be obtained by acting on the basic geodesic (or any other image) for any value of the angular variable. Therefore the correlator obtained from geodesics approximation can be written as a single sum (\ref{geodesicsorbifold}). It coincides with the GKPW expression and is smooth in the entire living space. 
 \begin{figure}[t]
\centering
\includegraphics[width=7cm]{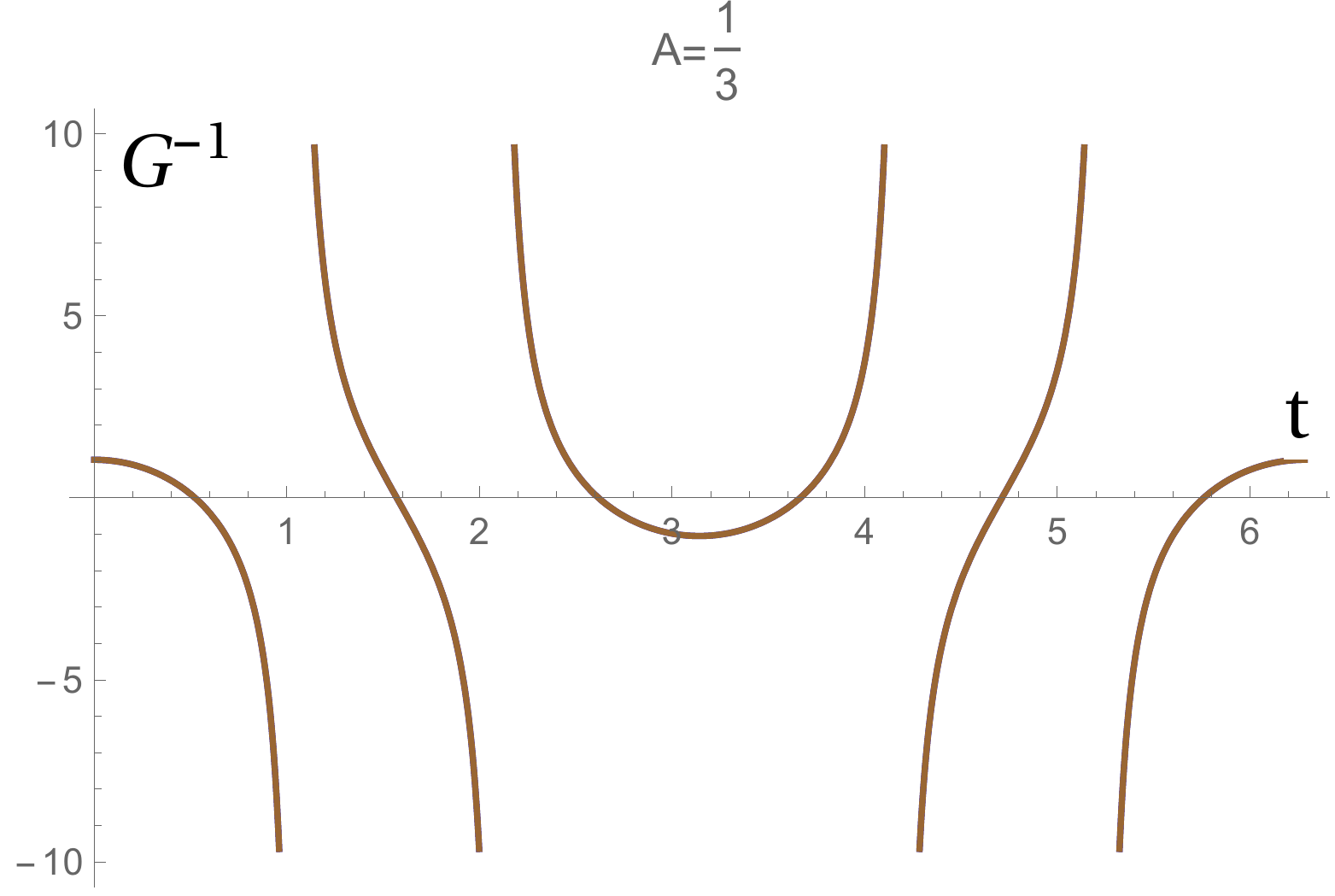}A.
\includegraphics[width=7cm]{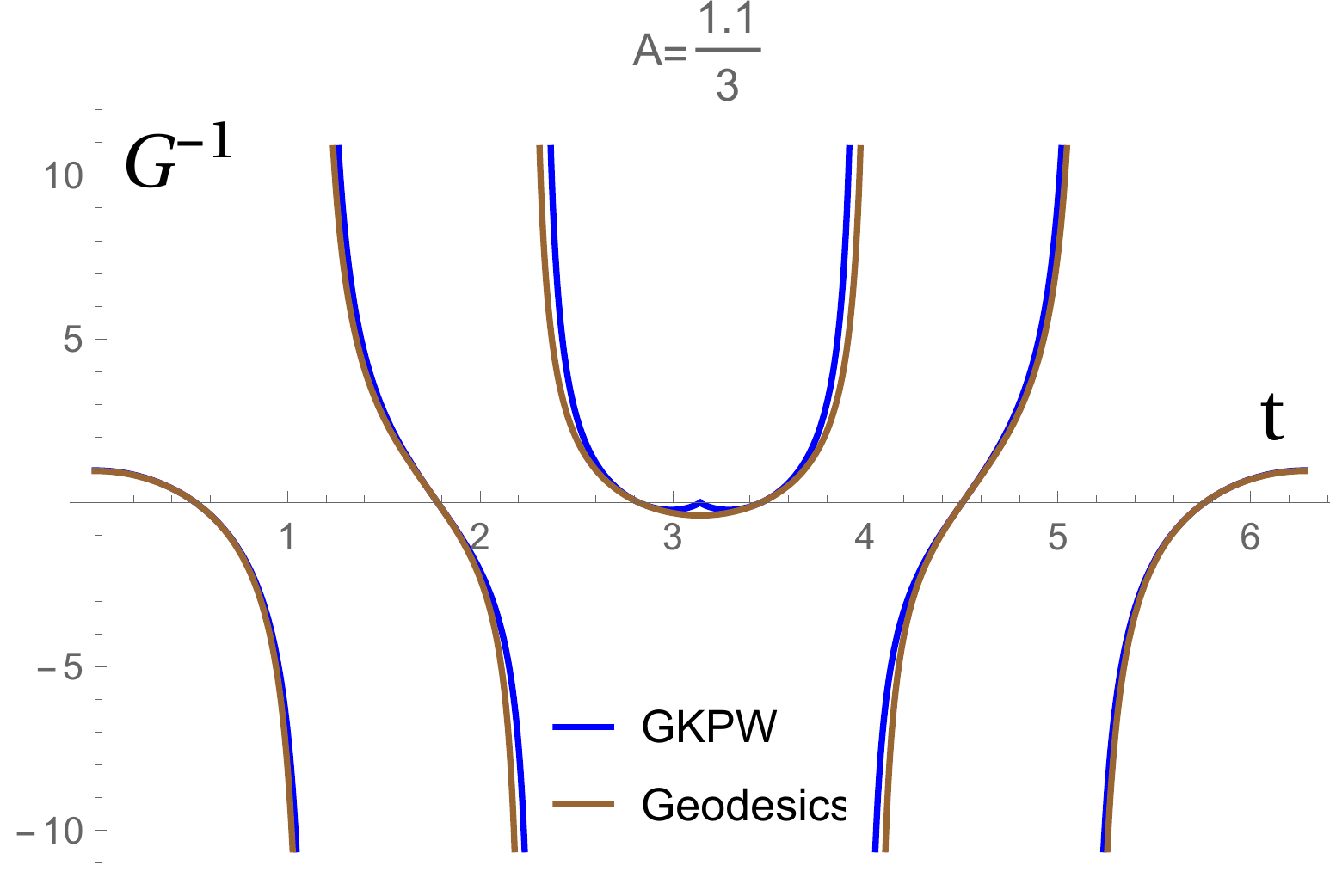}B.\\
$\,\,\,\,\,\,\,\,\,,\,\,\,$\\
\includegraphics[width=7cm]{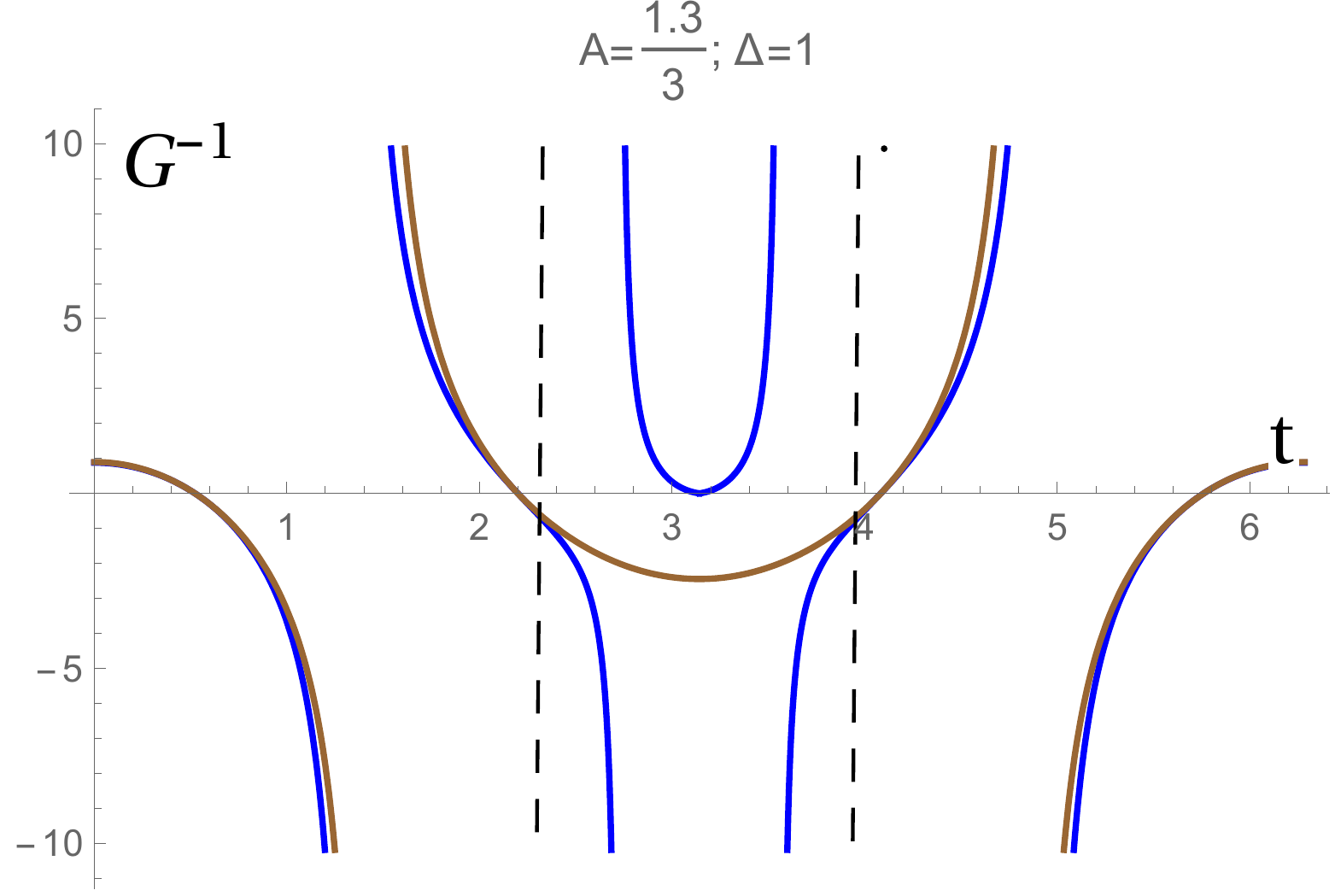}C.
\includegraphics[width=7cm]{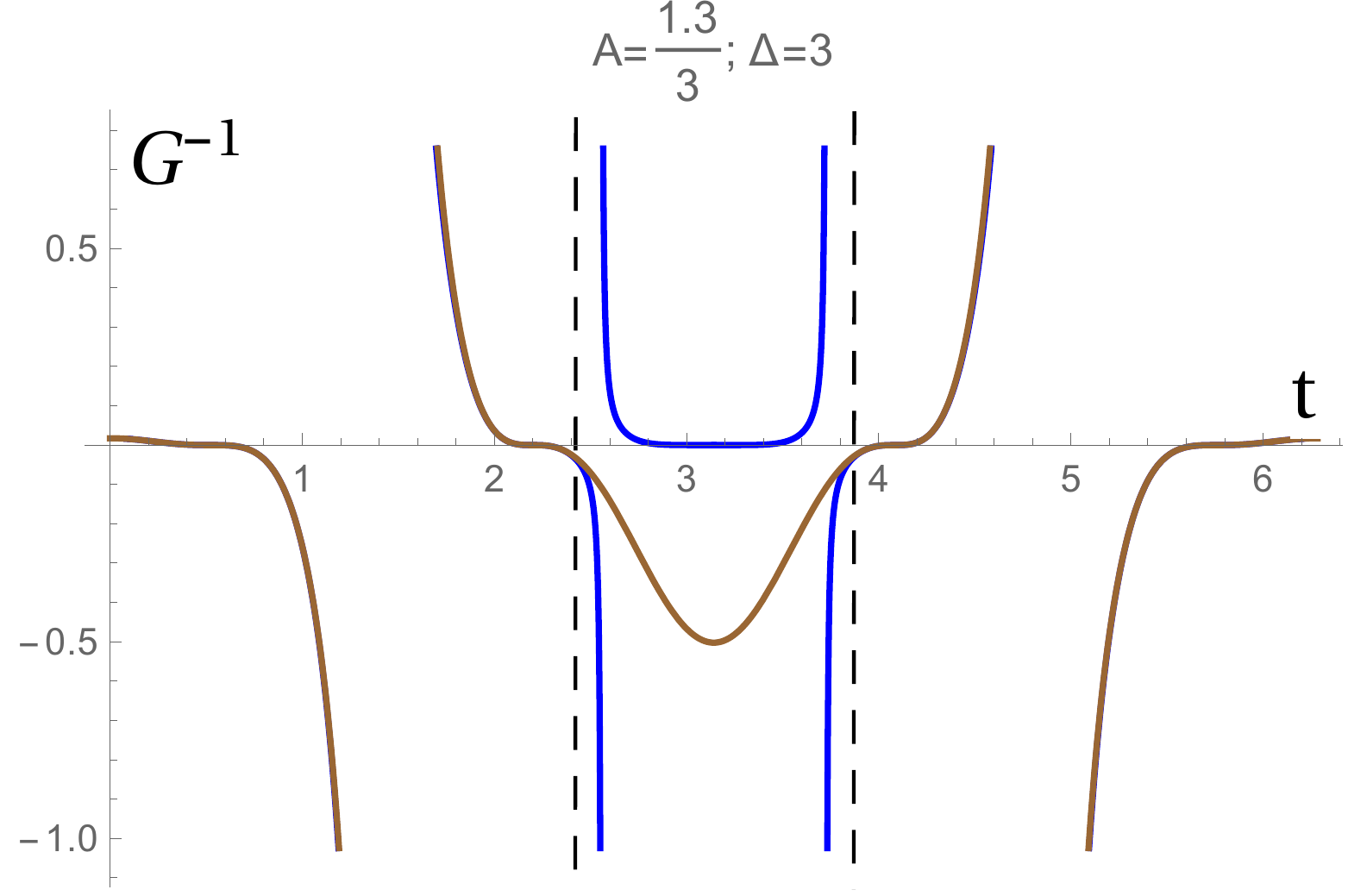}D.
\caption{Time dependence of the inverse correlators obtained by GKPW and geodesic prescriptions. Plots A-C show the increase of discrepancy between two prescriptions in case of $\Delta=1$ when the deficit parameter is close to the orbifold value $A=\frac13$. Plot D shows the discrepancy for $\Delta=3$. The value of angular variable is fixed $\vartheta = \frac{\pi}{6}$. 
} \label{Fig:time-dep}
\end{figure}
The answer for the inverse equal time correlator compared with the GKPW expression given by (\ref{GKPW-cone}) is shown in Fig. \ref{Fig:nu0-5large}. We see that contributions of discontinuities also diminish with the increase of the conformal weight, but the sign of corrections to the geodesic approximation is opposite to the small angle case, the GKPW value of the inverse correlator is sightly lower than that of the geodesic expression, and correction contributions in the long range region are much smaller than in the small deficit case. 
\subsection{Non-equal time correlators}
Here we examine the differences between the time dependencies of the GKPW answer (\ref{corrcyl})-(\ref{corrcylnu}) and the geodesic expression (\ref{geodesicsImages}). In Fig.\ref{Fig:time-dep} A-C we trace the increase of discrepancy between the two prescriptions when we slightly increase the value of the deficit parameter starting from $A=\frac13$. In this point the two prescriptions coincide:
$$\frac{\sin 3t}{\sin t}\frac{1}{\cos 3t-\cos 3\vartheta}$$\be
= \frac13 \left( \frac{1}{\cos t - \cos \vartheta} + \frac{1}{\cos t - \cos \left(\vartheta+\frac{2 \pi}{3}\right)} + \frac{1}{\cos t - \cos\left(\vartheta+\frac{4 \pi}{3}\right)} \right).
\ee
As we begin to deform the orbifold, we observe that some differences in the analytic structure of GKPW and geodesic expressions start to evolve.

First, consider zeros of the reverse correlator. These correspond to the singularities in the correlator itself. The geodesic correlator (\ref{geodesicsImages}) has singularities at lightcones corresponding to image points, and GKPW expressions (\ref{corrcyl}-\ref{corrcylnu}) obtain their singularities from the $(\cos t/A - \cos \vartheta/A)^{-1}$ and its derivatives in case of higher weights. In the general case these two sets do not coincide, but in the orbifold case one can derive trigonometric formulae similar to the above. We see in Fig. \ref{Fig:time-dep}B that when we move from the orbifold point, the GKPW reverse correlator obtains an additional zero. This happens because the change in the deficit parameter was too small to eliminate or add a term in the geodesic expression, but it was enough to have an impact on the denominator of the GKPW formula. This effect does not depend on the value of angle and, in general, on the number of geodesic images: we see in Fig.\ref{Fig:time-dep}C that the number of images has decreased, and GKPW expression reflects this as well, but still keeps its extra zero.

Second effect is related to the singularities of the reverse correlator, or zeros of the correlator itself. For GKPW expression, these come from the factor $\sin \frac{t}{A}$ in (\ref{corrcyl}) (and, again, its derivatives for higher weights in (\ref{corrcylnu})). In the geodesic prescription (\ref{geodesicsImages}) these come from the sum of all image denominators if one tries to bring the sum to a common denominator. Again, in the general case these two sets are different. In Fig.\ref{Fig:time-dep}C we see that when the number of terms in the sum is lowered by the constraint (\ref{geodesicsNumber}), the number of singularities of the reverse correlator decreases as well - in other words, the number of singularities is basically equal to the number of image geodesics, and thus to the number of lightcone zeros (on the interval $t \in [0, \pi]$). This is, however, not the case for the GKPW expression - the number of singularities is still the same, whereas the number of zeros has decreased compared to the Fig.\ref{Fig:time-dep}B case. 

This difference in the region between dashed lines between the geodesic and GKPW expressions is similar in its nature to the long-range contributions in the angular dependence of correlators discussed above. The comparison of plots C and D in Fig.\ref{Fig:time-dep}, that show the cases of different conformal dimensions at $A=\frac{1.3}{3}$, illustrates that the increase of the conformal dimension makes the geodesic prescription approach the GKPW expression where it has an extra zero, but the singularities between the dashed lines for certain values of the deficit parameters are still unique to the GKPW expression for general $\Delta$. Thus, unlike the long-range equal time case, the analytic structure of the reverse GKPW expression is not completely reproduced even in the large $\Delta$ limit. 

\section {Conclusion}

We have calculated the two-point boundary correlator in the AdS space with a static conical defect using the GKPW prescription for a scalar field. The version of the holographic prescription that we have used is formulated in the Lorentzian signature and is based on the deformation of the temporal integration contour in the bulk partition function. It does not require continuation from the Euclidean case. Generally speaking, the resulting correlator does not retain the conformal symmetry on the boundary. 

In the case of integer deficit, when the angle deficit equals to $\gamma = \frac{2\pi}{r}$, $r \in \mathbb{Z}$, we have shown that two-point correlator is equal to the CFT correlator. It can be represented as a sum over images. Each image contribution is expressed in this case by an expression for the correlator in the empty AdS$_3$. 

Comparing these GKPW correlators with correlators obtained through geodesic approximation, we observe that in general case with increasing $\Delta$ the geodesic approximation reproduces the GKPW expression more precisely. However, for equal-time correlators we see that correlators obtained via the geodesic approximation exhibit non-trivial generally discontinuous behaviour in the region of large spatial separations, which significantly differs from the behaviour of the GKPW correlators. The situation is slightly different for cases of small and large defects.
\begin{itemize}
\item For small angle deficit, $\frac12<A<1$, we observe that the geodesic correlator exhibits discontinuities between regions of contributions of image and basic geodesic. We see that the value of the reverse correlator is significantly lower around the point $\vartheta=\pi A$ than that of the GKPW expression. The latter depends on angles continuously everywhere in the living space. 
\item For large angle deficit, $A<\frac12$, we observe that the geodesic correlator exhibits discontinuities between regions of contributions of different sets of images. Contrary to the previous case, we see that the value of the reverse correlator is higher around the point $\vartheta=\pi A$ than that of the GKPW expression.
\end{itemize}
In general, we see that long-range corrections have higher impact in the spacetimes with small deficit angles. 

We also have examined the temporal behaviour of correlators obtained from GKPW and geodesic prescriptions, and we see that there is a difference in temporal dependence of the geodesic correlator and the GKPW one, which is similar to the large-separation discrepancy, in both cases of large and small deficits. Notice that the large $\Delta$ limit does not reproduce the GKPW result completely in some temporal regions. Indeed, the number of singularities in the GKPW expression is higher than the number of singularities in the geodesic correlator, which corresponds to the total number of geodesics involved in the images prescription, unless the deficit parameter is $A=\frac{1}{r}$, $r \in \mathbb{Z}$, i. e. the space is a $\mathbb{Z}_r$-orbifold. In that case, as we have observed, the geodesic approximation gives the exact answer for the CFT correlator, which coincides with the GKPW expression in this case as well, and the images method for calculating the Green's function in GKPW prescription coincides with the geodesic images prescription. 

The presence of non-trivial long-range corrections in the general case appears to be a common property of geodesic approximation in the various locally AdS backgrounds with broken asymptotic conformal symmetry. In order to try to give a physical interpretation to this fact, we recall from the discussion in \ref{geodesicApp} that in our case the geodesic approximation can be thought of as the leading order of the WKB approximation to the full GKPW expression, so we can interpret the discrepancy in the long-range region as inapplicability of the
WKB approximation. The failure of the WKB approximation could be related with
 the essential role of quantum corrections to the geodesics correlators for longest geodesics. This could take place because in that region the longest-wave excitations which give significant contribution to the path integral (\ref{path}) near its saddle points begin to interact in a purely quantum way with the static particle located at the origin (recall that the longest geodesics in AdS$_3$ pass closest to the origin). This interpretation suggests that the geodesic prescription can exhibit distinction from GKPW expression for other multi-connected locally AdS backgrounds. Another interesting point is that the long-range region the saddle points of the path integral (\ref{path}) are closest to each other (as illustrated, for example, in (\ref{largeAgeodesic}) - for long-range correlations the lengths of basic and image geodesics are closest to each other). This hints us that a viable improvement of the geodesic prescription on multi-connected spaces perhaps could be constructed by accounting for quantum corrections from adjacent saddles in every term of the right hand side in (\ref{path}), and it is plausible that those corrections could be interpreted as the effect of quantum scattering of particles on the defect. We hope to obtain some further insight on this issue and its connection to the conformal symmetry breaking in the future studies.

\section*{Acknowledgements}

The authors are grateful to Dmitrii Ageev and Andrey Bagrov for useful discussions.
This work is supported by the Russian Science Foundation (project 14-50-00005, Steklov
Mathematical Institute).

  \end{document}